\documentclass[twocolumn,pra,showpacs,amsmath]{revtex4}

\usepackage{graphicx}

\begin{document}

\title
     {Quantum optics with quantum gases:

     controlled state reduction by designed light scattering}
\date{\today}

\author{Igor B. Mekhov}
\email{Igor.Mekhov@uibk.ac.at}
\author{Helmut Ritsch}
\affiliation{Institut f\"ur Theoretische Physik, Universit\"at
Innsbruck, Innsbruck, Austria}

\begin{abstract}
Cavity enhanced light scattering off an ultracold gas in an optical
lattice constitutes a quantum measurement with a controllable form
of the measurement back-action. Time-resolved counting of scattered
photons alters the state of the atoms without particle loss
implementing a quantum nondemolition (QND) measurement. The
conditional dynamics is given by the interplay between
photodetection events (quantum jumps) and no-count processes. The
class of emerging atomic many-body states can be chosen via the
optical geometry and light frequencies. Light detection along the
angle of a diffraction maximum (Bragg angle) creates an atom-number
squeezed state, while light detection at diffraction minima leads to
the macroscopic superposition states (Schr{\"o}dinger cat states) of
different atom numbers in the cavity mode. A measurement of the
cavity transmission intensity can lead to atom-number squeezed or
macroscopic superposition states depending on its outcome. We
analyze the robustness of the superposition with respect to missed
counts and find that a transmission measurement yields more robust
and controllable superposition states than the ones obtained by
scattering at a diffraction minimum.

\end{abstract}


\maketitle

\section{Introduction}

Both, quantum optics and physics of ultracold quantum gases represent
nowadays the well-established and actively developing fields of
modern quantum science \cite{BlochDalibard,Lewenstein}. However, the
interaction between the two fields is far from being close.

Historically, classical optics treating the light as classical
electromagnetic waves has become one of the most developed and
fruitful fields of physics. It has provided us a lot of
technological breakthroughs, e.g., the highest level of measurement
precision. Quantum optics, which considers the light as quantum
particles (photons), thus going beyond the mean-field (classical)
description of the electromagnetic waves, currently is also a
well-developed field, both theoretically and experimentally
\cite{Scully}.

The progress in laser cooling techniques in the last decades of the
20th century led to the foundation of a new field of atom physics:
atom optics. It was shown that the matter waves of ultracold atoms
can be treated similar to light waves in classical optics and can be
manipulated using the forces and potentials of laser light beams.
The quantum properties of matter waves beyond the mean-field
description became accessible after 1995, when the first
Bose-Einstein condensate (BEC) and many other fascination quantum
states of bosonic and fermionic ultracold atoms were obtained
\cite{BlochDalibard,Lewenstein}. An exciting demonstration of
"quantum atom optics" was presented in 2002, when the quantum phase
transition between two states of atoms with nearly the same mean
density, but radically different quantum fluctuations was obtained:
superfluid (SF) to Mott insulator (MI) state transition
\cite{Jaksch,BlochSFMI}.

The roles of light and matter in optics and atom optics are
completely reversed. Various devices known in optics as
beam-splitters, mirrors, diffraction grating, etc. are created using
light forces and applied for matter waves. However, up to now, the
absolute majority of even very involved setups and theoretical
models of quantum atom optics treat light as an essentially
classical axillary tool to prepare and probe intriguing atomic
states. In this contexts, the periodic micropotentials of light
(optical lattices) play the role of cavities in optics enabling one to
store and manipulate various atomic quantum states.

Quantum optics with quantum gases should close the gap between
quantum optics and atom optics by addressing phenomena, where the
quantum natures of both light and matter play equally important
role. Experimentally, such an ultimate quantum level of the
light-matter interaction became feasible only recently, when the
quantum gas was coupled to the mode of a high-Q cavity
\cite{Brennecke,Colombe,Slama,Science08}. Even early theoretical
works on scattering of quantized light from a BEC was not realized
so far \cite{moore,pu,you,idziaszek,mustPRA62,mustPRA64,
javPRL,javPRA,ciracPRL,ciracPRA,saito,prataviera,javOL}. However, it
is cavity quantum electrodynamics (QED) with quantum gases that will
provide the best interplay between the atom- and light-stimulated
quantum effects.

On the one hand, the quantum properties of atoms will manifest
themselves in the scattered light, which will lead to novel
nondestructive methods of probing and manipulating atomic states by
light measurement
\cite{PRL07,NP,PRA07,PRL09,LP09,Meystre07,Meystre09,PolzikNP,BhatOC,RuostPRA08,ZhangPRA08,Cirac1,GuoPRA09}.
On the other hand, the quantization of light (i.e. trapping
potentials) will modify atomic manybody dynamics well-known only for
classical potentials and give rise to novel quantum phases
\cite{PRL05,EPJD08,LewPRL09,LewNJP,AndrNJP,AndrPRA09,DomEPJD,DomPRL}.
This paper addresses the first problem.

In this work, we will extend our treatment of the detection of light
scattered from ultracold atoms in optical lattices presented in Ref.
\cite{PRL09}, where the quantum measurement of light was considered
as a method to prepare particular quantum states of atoms thanks to
the measurement back-action. Except detailing the theoretical
approach, we will consider a different optical configuration of the
measurement process: cavity transmission spectroscopy instead of
light scattering. As we will show, it turns out, that such a
configuration will allow us much better control and flexibility in
atomic state preparation. This is in particular true for the
preparation of macroscopic superposition states (known as
Schr{\"o}dinger cat states), which can be more robust in comparison
to those considered before \cite{PRL09,AndrNJP}.

We will show that detecting the light scattered from ultracold atoms
in an optical lattice enables one to prepare various types of the
atom-number squeezed and macroscopic superposition states. An
important point, is that the type of the state and its properties
strongly depend on the optical geometry. Varying the optical
parameters (angles between the laser beams and lattice atoms, light
frequencies, or lattice period) one can prepare various quantum
states of ultracold atoms. Moreover, as the optical measurement is
nondestructive, in the sense of the quantum nondemolition
measurements (QND), one can make sequential measurements on the same
sample without completely destroying its quantum properties.

Generalizing the methods developed for spin squeezing in thermal
atomic ensembles
\cite{PolzikHot,Holland,Genes,MolmerPRA08,MolmerPRA09} for the case
of light scattering from ultracold quantum gases will enable ones to
reach the unprecedented level of the measurement precision, which is
required, e.g., for atomic clocks and the detection of gravitational
waves.

The paper is organized as follows. In Sec. II, a general theoretical
model is formulated, which, in Sec. III, is extended to the quantum
back-action of the light measurement on the atomic state. Section IV
is devoted to the scattering of the transverse probe in to the
cavity. In Sec. V, the properties of photon statistics are analyzed.
The state preparation by the cavity transmission measurement is
discussed in Sec. VI. The robustness of the macroscopic
superposition states is analyzed in Sec. VII. The main results are
summarized in Sec. VIII.

\section{General model}

\begin{figure}
\scalebox{0.45}[0.45]{\includegraphics{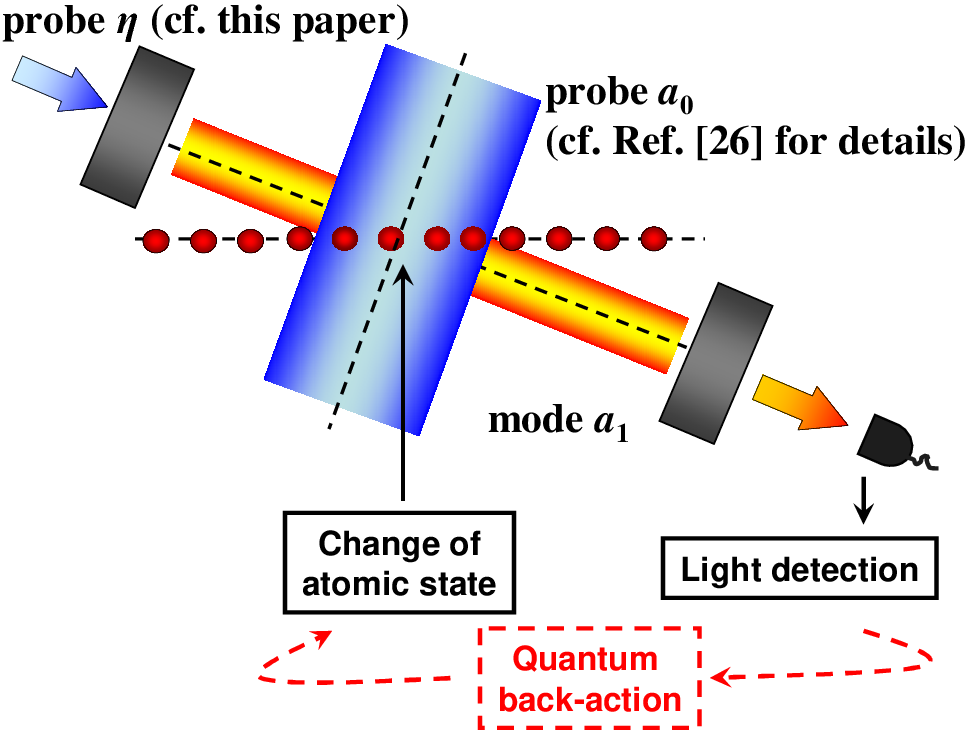}}
\caption{\label{fig1}(Color online) Setup. A lattice is illuminated
by the transverse probe $a_0$ and probe through a mirror $\eta$. The
photodetector measures photons leaking the cavity. Due to the
quantum back-action, the light measurement leads to the modification
of the atomic quantum state.}
\end{figure}

We consider $N$ ultracold atoms trapped in a periodic optical
lattice potential with $M$ sites. In addition to the strong trapping
beams, a subset of $K$ sites, with $K\le M$, is illuminated by a
weak probe at some angle to the lattice atoms. The scattered light
is collected by a cavity. The photons leaking out of the cavity are
counted then with a photodetector. The schematic setup is shown in
Fig. 1, where, for simplicity, the trapping beams are not shown. In
general, the setup is not restricted by a simplified scheme in Fig.
1: the lattice can be one-, two-, or three-dimensional; the probe
and cavity modes can be standing or traveling waves, or they can be
even formed by two different modes of the same cavity. $K$
illuminated sites can be selected in some nontrivial way, e.g., each
second lattice site can be illuminated by choosing the probe
wavelength two times larger than the lattice period. In addition to
the external probe, a probing through a cavity mirror will be
considered.

The main goal of this paper is to study how the measurement of the
photons leaking out of the cavity will affect the quantum state of
the ultracold atoms. Such a measurement back-action is possible due
to the entanglement between light and matter, which develops during
the interaction process. According to quantum mechanics, in the
presence of the entanglement, the measurement of one of the quantum
subsystems (light) will also affect another quantum subsystem
(atoms).

We will use the physical model presented in details in Ref.
\cite{PRA07}. The theoretical formulation starts with a generalized
Bose-Hubbard model including the quantization of light. As here,
unlike Ref. \cite{Hashem}, the far off-resonant light scattering
will be considered, the role of the atomic excited states is not
important and they can be adiabatically eliminated from the
dynamics. As we will be focused here on the quantum measurement
process, it is reasonable to neglect several processes have been
considered in other works \cite{EPJD08,AndrNJP,AndrPRA09} to avoid
the complications. One of such processes to be neglected is the
details of atomic dynamics. The tunneling of atoms in the lattice
potential plays indeed an important role in establishing a
particular quantum atomic state. However, after the state is
established, one can assume that the scattering of the probe occurs
on the time-scale faster than slow tunneling. Thus, from the light
scattering point of view, the atomic distribution can be considered
frozen and the tunneling is not important. It is especially
reasonable to neglect the tunneling dynamics in this paper, because
we will show that even after that rather obvious dynamics is
neglected, there is still non-trivial dynamics (quantum jumps)
exclusively associated with the quantum measurement process, which
is the main subject of our present work.

After those simplifications, the Hamiltonian of the problem takes
the form:
\begin{eqnarray}\label{1}
H=\hbar(\omega_1 + U_{11} \hat{D}_{11}) a^\dag_1 a_1+
\hbar U_{10}(\hat{D}^*_{10}a^*_0a_1 + \hat{D}_{10}a_0a^\dag_1)  \nonumber \\
-i\hbar(\eta^* a_1 - \eta a^\dag_1),
\end{eqnarray}
where $a_1$ is the cavity-mode annihilation operator of the
frequency $\omega_1$. The external probe of the frequency $\omega_p$
is assumed to be in a coherent state, thus its amplitude is given by
a c-number $a_0$. The spatial mode functions of the probe and cavity
modes are $u_{1,0}({\bf r})$. $U_{lm}=g_lg_m/\Delta_a$ ($l,m=0,1$),
where $g_{1,0}$ are the atom-light coupling constants,
$\Delta_a=\omega_1 - \omega_a$ is the cavity-atom detuning, $\eta$
is the amplitude of the additional probing through a mirror at the
frequency $\omega_p$. We assumed the probe-cavity detuning
$\Delta_{p}=\omega_{p}-\omega_1 \ll\Delta_a$. The operators
$\hat{D}_{lm}= \sum_{j=1}^K{u_l^*({\bf r}_j)u_m({\bf
r}_j)\hat{n}_j}$ sum contributions from all illuminated sites with
the atom-number operators $\hat{n}_j$ at the position ${\bf r}_j$.

The first term in Eq.~(\ref{1}) describes the atom-induced shift of
the cavity resonance. The second one reflects scattering
(diffraction) of the probe $a_0$ into a cavity mode $a_1$. For a
quantum gas the frequency shift and probe-cavity coupling
coefficient are operators, which leads to different light scattering
from various atomic quantum states~\cite{PRL07,NP,PRA07,LP09}.

The Hamiltonian (\ref{1}) describes QND measurements of the
variables related to $\hat{D}_{lm}$ measuring the photon number
$a^\dag_1a_1$, as the criteria for the QND measurements are
fulfilled \cite{Brune}. Note, that one has a QND access to various
many-body variables, as $\hat{D}_{lm}$ strongly depend on the
lattice and light geometry via $u_{0,1}({\bf r})$. This is an
advantage of the lattice comparing to single- or double-well setups,
where the photon measurement back-action was considered
\cite{PRL12a,PRL12b,PRL12c,PRL12d}. Moreover, such a geometrical
approach can be extended to other quantum arrays, e.g., ion strings
\cite{ions}.

For example, one can consider a 1D lattice of the period $d$ with
atoms trapped at $x_j=jd$ ($j=1,2,..,M$). In this case, the
geometric mode functions can be expressed as follows: $u_{0,1}({\bf
r}_j)=\exp (ijk_{0,1x}d+\phi_{0,1j})$ for traveling waves, and
$u_{0,1}({\bf r}_j)=\cos (jk_{0,1x}d +\phi_{0,1j})$ for standing
waves, where $k_{0,1x}=|{\bf k}_{0,1}|\sin\theta_{0,1}$,
$\theta_{0,1}$ are the angles between mode wave vectors ${\bf
k}_{0,1}$ and a vector normal to the lattice axis. In the plane-wave
approximation, additional phases $\phi_{0,1j}$ are $j$-independent.
The general angular distributions of light scattered from ultracold
atoms in optical lattices were presented in
Refs.~\cite{PRL07,PRA07,LP09}

For some geometries, $\hat{D}_{11}$ can reduce to the operator
$\hat{N}_K=\sum_{j=1}^K\hat{n}_j$ of the atom number at $K$ sites
\cite{PRL07,NP,PRA07} (if $a_1$ is a traveling wave at an arbitrary
angle to the lattice, or the standing wave with atoms trapped at the
antinodes). If the probe and cavity modes are coupled at a
diffraction maximum (Bragg angle), i.e., all atoms scatter light in
phase, $u_1^*({\bf r}_j)u_0({\bf r}_j)=1$, the probe-cavity coupling
is maximized, $\hat{D}_{10}=\hat{N}_K$. If they are coupled at a
diffraction minimum, i.e., the neighboring atoms scatter out of
phase, $\hat{D}_{10}=\sum_{j=1}^K (-1)^{j+1}\hat{n}_j$ is the
operator of number difference between odd and even sites. Thus, the
atom number as well as number difference can be nondestructively
measured. Note, that those are just two of many examples of how a
QND-variable, and thus the projected state, can be chosen by the
geometry.

\section{Measurement back-action}

In this section we present the solution for the quantum state of the
coupled light-matter system including the measurement process.
Importantly, it is possible to obtain an analytical solution with a
very transparent physical meaning thanks to the approximations used
(slow tunneling and coherent state of the external probes).

The initial motional state of the ultracold atoms trapped in the
periodic lattice potential at the time moment $t=0$ can be
represented as

\begin{eqnarray}\label{2}
|\Psi^a(0)\rangle =\sum_{q}c_q^0 |q_1,..,q_M\rangle,
\end{eqnarray}
which is a quantum superposition of the Fock states corresponding to
all possible classical configurations $q=\{q_1,..,q_M\}$ of $N$
atoms at $M$ sites, where $q_j$ is the atom number at the site $j$.
For each classical configuration $q$, the total atom number is
conserved: $\sum_j^M q_j=N$. This superposition displays the
uncertainty principle, stating that at ultralow temperatures even a
single atom can be delocalized in space, i.e., there is a
probability to find an atom at any lattice site. We will show, how
this atomic uncertainty is influenced by the light detection.

For example, for a limiting case of the MI state, where the atom
numbers at each lattice site are precisely known, only one Fock
state will exist in Eq.~(\ref{2}): $|\Psi_\text{MI}\rangle
=|1,1,..,1\rangle$ for the MI with one atom at each site. On the
other hand, the SF state is given by the superposition of all
possible classical configurations with multinomial coefficients:

\begin{eqnarray}\label{3}
|\Psi^a_\text{SF}\rangle=\frac{1}{(\sqrt{M})^N}\sum_{q}
\sqrt{\frac{N!}{q_1!q_2!...q_M!}}|q_1,q_2,..q_M\rangle.
\end{eqnarray}
Thus, the atom number at a single site as well as the atom number at
$K<M$ sites are uncertain in the SF state.

As an initial condition, we assume that at the time moment $t=0$ the
light and matter are disentangled, and the initial state of light is
a coherent state with the amplitude $\alpha_0$. Thus, the initial
quantum state of the system is given by the product state
$|\Psi(0)\rangle=|\Psi^a(0)\rangle |\alpha_0\rangle$. In particular,
initially, the light can be in the vacuum state $|0\rangle$.

We use the open system approach \cite{Carmichael} to describe the
continuous counting of the photons leaking out the cavity of the
cavity decay rate $\kappa$. According to that approach, when the
photon is detected at the moment $t_i$, the quantum jump occurs, and
the state instantaneously changes to a new one obtained by applying
the cavity photon annihilation operator $|\Psi_c(t_i)\rangle
\rightarrow a_1|\Psi_c(t_i)\rangle$ and renormalization (the
subscript $c$ underlines that we deal with the state conditioned on
the photocount event). Between the photocounts, the system evolves
with a non-Hermitian Hamiltonian $H-i\hbar\kappa a^\dag_1a_1$. Such
an evolution gives a quantum trajectory for $|\Psi_c(t)\rangle$
conditioned on the detection of photons at times $t_1,t_2,...$. The
probability of the photon escape within the time interval $t$ is
$2\kappa t\langle a^\dag_1a_1\rangle_c$, where $\langle
a^\dag_1a_1\rangle_c$ is the conditional photon number in the
cavity, i.e., the photon number calculated for the conditional
quantum state $|\Psi_c(t)\rangle$.

The state $|\Psi_c(t)\rangle$ should be found by solving the
Schr{\"o}dinger equation with the non-Hermitian Hamiltonian for
no-count intervals and applying the jump operator $a_1$ at the
moments of photocounts. Thanks to the slow tunneling approximation,
the Hamiltonian (\ref{1}) does not mix the Fock states in the
expression (\ref{2}). So, the problem is significantly reduced to
separate finding solutions for each classical atomic configuration
$q=\{q_1,..,q_M\}$, after that the full solution will be given by
the superposition of those solutions.

The next simplification appears thanks to the use of the external
probes in the coherent state. It is
known~\cite{Denis,GardinerZoller} that, if a coherent probe
illuminates a classical atomic configuration $q$ in a cavity, the
light remains in a coherent state, however, with some pre-factor:
$\exp[\Phi_q(t)]|\alpha_q(t)\rangle$. The pre-factor is indeed not
important for a single classical configuration as it disappears
after the renormalization, but it will play a role, when the
superposition of classical solutions with different pre-factors will
be considered. Moreover, the light amplitude $\alpha_q(t)$ is simply
given by the solution of a classical Maxwell's equation:
\begin{eqnarray}\label{4}
\alpha_q(t)=\frac{\tilde{\eta}-iU_{10}\tilde{a}_0D^q_{10}}{i(U_{11}
D^q_{11}-\Delta_p)+\kappa}e^{-i\omega_pt}+ \nonumber \\
\left(\alpha_0 - \frac{\tilde{\eta}-iU_{10}
\tilde{a}_0D^q_{10}}{i(U_{11}
D^q_{11}-\Delta_p)+\kappa}\right)e^{-i(\omega_1+U_{11}D^q_{11})t-\kappa
t},
\end{eqnarray}
where we introduced the constant probe amplitudes $\tilde{a}_0$ and
$\tilde{\eta}$ as $a_0=\tilde{a}_0\exp{(-i\omega_pt)}$ and
$\eta=\tilde{\eta}\exp{(-i\omega_pt)}$; $D^q_{lm}=
\sum_{j=1}^K{u_l^*({\bf r}_j)u_m({\bf r}_j)q_j}$ is a realization of
the operator $\hat{D}_{lm}$ at the configuration $q=\{q_1,..q_M\}$.
As in classical optics, the first term in Eq.~(\ref{4}) gives the
oscillations at the probe frequency, while the second term gives the
transient process with the oscillations at the cavity frequency
shifted by the dispersion, $\omega_1+U_{11}D^q_{11}$, which decays
with the rate $\kappa$. In the following, we introduce the slowly
varying light amplitude $\tilde{\alpha}_q(t)$ as
$\alpha_q(t)=\tilde{\alpha}_q(t)\exp{(-i\omega_pt)}$ and, for the
notation simplicity, will drop the tilde sign in all amplitudes
$\tilde{a}_0$, $\tilde{\eta}$, and $\tilde{\alpha}_q(t)$.

The function $\Phi_q(t)$ in the pre-factor is a complex one and is
given by
\begin{eqnarray}\label{5}
\Phi_q(t)=\int_0^t\left[\frac{1}{2}(\eta\alpha^*_q-iU_{10}
a_0D^q_{10}\alpha^*_q-\text{c.c.})-\kappa|\alpha_q|^2\right]dt,
\end{eqnarray}
where the light amplitude $\alpha_q(t)$ is given by Eq.~(\ref{4}).

First, let us consider the solution for the atomic state initially
containing a single Fock state $|q_1,..,q_M\rangle$. The solution
for light is given by the solution for the classical configuration
$q=\{q_1,..,q_M\}$. So, the evolution is given by the product state
$|q_1,..,q_M\rangle|\alpha_q(t)\rangle$. An important property of
this solution is that a quantum jump does not change the state,
since applying the jump operator $a_1$ simply leads to the
pre-factor $\alpha_q(t)$, which disappears after the
renormalization. Therefore, even in the presence of photocounts
(i.e. quantum jumps), the time evolution is continuous and is given
by Eq.~(\ref{4}): after a transient process for $t<1/\kappa$, the
steady state for $\alpha_q(t)$ is achieved. Note, that in contrast
to many problems in quantum optics \cite{Carmichael}, where the
steady state is a result of averaging over many quantum trajectory,
here, the steady state appears even at a single quantum trajectory
for $t>1/\kappa$. This is a particular property of the coherent
quantum state. The continuity of evolution of the state
$|q_1,..,q_M\rangle|\alpha_q(t)\rangle$, and, hence, more generally,
the unnormalized state including the pre-factor,
$\exp[\Phi_q(t)]|q_1,..,q_M\rangle|\alpha_q(t)\rangle$,
independently of the presence of the quantum jumps provides us a
significant mathematical simplification. Moreover, we will use the
result that after the time $t>1/\kappa$, all light amplitudes
$\alpha_q$ are constant for all Fock states $|q_1,..,q_M\rangle$.

Let us now consider the full initial state given by the
superposition (\ref{2}). As stated, the evolution of each term is
independent and contains the continuous part
$\exp[\Phi_q(t)]|q_1,..,q_M\rangle|\alpha_q(t)\rangle$. Thus,
applying the jump operators at the times of the photodetections
$t_1,t_2,..t_m$ leads to the following analytical solution for the
conditional quantum state at the time $t$ after $m$ photocounts:
\begin{eqnarray}\label{6}
|\Psi_c(m,t)\rangle
=\frac{1}{F(t)}\sum_{q}\alpha_q(t_1)\alpha_q(t_2)...\alpha_q(t_m)
 \nonumber\\
\times  e^{\Phi_q(t)}c_q^0|q_1,...,q_M\rangle|\alpha_q(t)\rangle,
\end{eqnarray}
where
\begin{eqnarray}
F(t)=
\sqrt{\sum_{q}|\alpha_q(t_1)|^2|\alpha_q(t_2)|^2...|\alpha_q(t_m)|^2
|e^{\Phi_q(t)}|^2|c_q^0|^2} \nonumber
\end{eqnarray}
is the normalization coefficient.

In contrast to a single atomic Fock state, the solution (\ref{6}),
in general, is not factorizable into a product of the atomic and
light states. Thus, in general, the light and matter are entangled.
Moreover, in contrast to a single Fock state, the quantum jump
(applying $a_1$) changes the state, and the evolution of the full
$|\Psi_c(m,t)\rangle$ is not continuous.

The general solution (\ref{6}) valid for all times simplifies
significantly for $t>1/\kappa$, when all $\alpha_q$ are constants,
and if the first photocount occurred at $t_1>1/\kappa$, when all
$\alpha_q$ has already become constants. The latter assumption is
especially probable for the small cavity photon number, since the
probability of the photon escape within the time interval $t$ is
$2\kappa t\langle a^\dag_1a_1\rangle_c$. This solution takes the
form
\begin{eqnarray}
|\Psi_c(m,t)\rangle =\frac{1}{F(t)}\sum_{q}\alpha_q^m e^{\Phi_q(t)}
c_q^0 |q_1,...,q_M\rangle|\alpha_q\rangle, \label{7}\\
\alpha_q=\frac{\eta-iU_{10} a_0D^q_{10}}{i(U_{11}
D^q_{11}-\Delta_p)+\kappa}, \label{8}\\
\Phi_q(t)=-|\alpha_q|^2\kappa t+(\eta\alpha^*_q-iU_{10}
a_0D^q_{10}\alpha^*_q-\text{c.c.})t/2, \label{9}
\end{eqnarray}
with the normalization coefficient
\begin{eqnarray}
F(t)= \sqrt{\sum_{q}|\alpha_q|^{2m}e^{-2|\alpha_q|^2\kappa
t}|c_q^0|^2}. \nonumber
\end{eqnarray}

The solution (\ref{7}) does not depend on the photocount times
$t_1,t_2,..t_m$ any more. Note however, that even for $t>1/\kappa$,
when all $\alpha_q$ reached their steady states and are constants,
the solution (\ref{7}) is still time-dependent. Thus, the time
$t=1/\kappa$ is not a characteristic time scale for the steady state
of the full solution (\ref{6}) and (\ref{7}). The stationary light
amplitudes $\alpha_q$ in (\ref{8}) are given by the Lorentz
functions in the absolute correspondence with classical optics. The
function $\Phi_q(t)$ has also simplified and contains the first real
term responsible for the amplitudes of the coefficients in the
quantum superposition, and the second imaginary term responsible for
their phases.

The solutions (\ref{6}) and (\ref{7}) show, how the probability to
find the atomic Fock state $|q_1,..,q_M\rangle$ (corresponding to
the classical configuration $q$) changes in time. Such a change in
the atomic quantum state appears essentially due to the measurement
of photons and is a direct consequence of the light-matter
entanglement: according to quantum mechanics, by measuring one of
the entangled subsystem (light) one also affects the state of
another subsystem (atoms). Now we can focus on that purely
measurement-base dynamics, since other obvious sources of the
time-evolution (e.g. tunneling) we neglected in our model. The
initial probability to find the Fock state $|q_1,..,q_M\rangle$ is
$p_q(0)=|c^0_q|^2$. From Eq.~(\ref{6}) the time evolution of this
probability is given by
\begin{eqnarray}\label{10}
p_q(m,t)= |\alpha_q(t_1)|^2|\alpha_q(t_2)|^2...|\alpha_q(t_m)|^2
 \\ \nonumber
 \times|e^{\Phi_q(t)}|^2p_q(0)/F^2(t).
\end{eqnarray}
For $t,t_1>1/\kappa$ [cf. Eq.~(\ref{7})] it reduces to
\begin{eqnarray}\label{11}
p_q(m,t)= |\alpha_q|^{2m}e^{-2|\alpha_q|^2\kappa t}p_q(0)/F^2(t).
\end{eqnarray}

In the following, we will demonstrate the applications of the
general solutions (\ref{6}) and (\ref{7}) in the examples, where,
for simplicity, only a single statistical quantity is important,
instead of the whole set of all possible configurations $q$. As
particular examples, we will consider the cases, where that
statistical quantity (let us now call it $z$) is the atom number at
$K$ lattice sites or the atom number difference between odd and even
sites. Thus, instead of the huge number of detailed probabilities
$p_q(m,t)$, we will be interested in the probability $p(z,m,t)$ to
find a particular value of $z$. For the initial state (\ref{2}) at
$t=0$, $p_0(z)=\sum_{q'} |c_{q'}^0|^2$ with the summation over all
configurations $q'$ having the same $z$. As under our assumptions
all light amplitudes $\alpha_q(t)$ depend only on $z$, we change
their subscript to $z$ and write the probability to find a
particular value of $z$ at time $t$ after $m$ photocounts:
\begin{eqnarray}\label{12}
p(z,m,t)= |\alpha_z(t_1)|^2|\alpha_z(t_2)|^2...|\alpha_z(t_m)|^2
 \\ \nonumber
 \times e^{2\text{Re}\Phi_z(t)}p_0(z)/F^2(t),
\end{eqnarray}
For $t,t_1>1/\kappa$ it reduces to
\begin{eqnarray}\label{13}
p(z,m,t)= |\alpha_z|^{2m}e^{-2|\alpha_z|^2\kappa
t}p_0(z)/F^2(t), \\
F(t)= \sqrt{\sum_{z}|\alpha_z|^{2m}e^{-2|\alpha_z|^2\kappa
t}p_0(z)}. \nonumber
\end{eqnarray}

In the following we will consider the solution (\ref{7}). When the
time progresses, both $m$ and $t$ increase with an essentially
probabilistic relation between them. The Quantum Monte Carlo method
\cite{Carmichael} establishes such a relation, thus giving a quantum
trajectory. Note, that thanks to the simple analytical solution
(\ref{7}), the method gets extremely simple. The evolution is split
into small time intervals $\delta t_i$. In each time step, the
conditional photon number is calculated in the state Eq.~(\ref{7}),
and the probability of the photocount within this time interval
$2\kappa \langle a^\dag_1a_1\rangle_c \delta t_i$ is compared with a
random number $0<\epsilon_i<1$ generated in advance, thus, deciding
whether the detection (if $2\kappa \langle a^\dag_1a_1\rangle_c
\delta t_i>\epsilon_i$) or no-count process (otherwise) has
happened.

\section{Transverse probing}

In this section we will consider a case, where only the transverse
probe $a_0$ is present, while the probe through the mirror does not
exist, $\eta=0$. We presented this situation in the previous letter
\cite{PRL09}. Here we remind the most important results and
underline the difficulties in the state preparation using the
transverse probing. In Sec. VI, we will switch to a new geometry,
probing through a mirror, and will show that such a geometry
contains the features similar to the transverse probing, but is more
flexible and enables us to solve the difficulties associated with
the transverse probing scheme.

In this section we neglect the dispersive frequency shift assuming
that $U_{11} D^q_{11}\ll\kappa$ or $\Delta_p$. Thus, the light
amplitudes $\alpha_q$ will only depend on the quantity $D^q_{10}$.
In Sec. VI, in contrast, we will focus on the case, where the
dispersive mode shift is very important.

\subsection{Preparation of the atom-number squeezed states}

The measurement of photons scattered in the direction of a
diffraction maximum (Bragg angle) leads to a preparation of a state
with the reduced (squeezed) fluctuations of the atom number at the
lattice region with $K$ illuminated sites \cite{PRL09}. The
condition of the diffraction maximum for the scattering of light
from the probe wave $a_0$ into the cavity mode $a_1$ is the
following: the atoms at all lattice sites scatter the light in phase
with each other. For the plain standing or traveling waves, this
condition means that in the expression for the operator
$\hat{D}_{10}= \sum_{j=1}^K{u_1^*({\bf r}_j)u_0({\bf
r}_j)\hat{n}_j}$, $u_1^*({\bf r}_j)u_0({\bf r}_j)=1$ for all sites
$j$. Thus, $\hat{D}_{10}=\hat{N}_K$ is reduced to the operator of
the atom number at $K$ illuminated sites. In Eq.~(\ref{7}), after
neglecting the dispersive frequency shift, the only statistical
quantity is $D^q_{10}$ giving the atom number at $K$ sites for the
configuration $q$. We will call this single statistical quantity as
$z$: $D^q_{10}=z$, which varies between 0 and $N$ reflecting all
possible realizations of the atom number at $K$ sites.

From Eq.~(\ref{8}), the light amplitudes in the diffraction maximum
are proportional to the atom number $z$:
\begin{eqnarray}\label{14}
\alpha_z=Cz, \text {with   }
C=\frac{iU_{10}a_0}{(i\Delta_p-\kappa)}.
\end{eqnarray}
Thus, the probability to find the atom number $z$ is given from
Eq.~(\ref{13}) by
\begin{eqnarray}\label{15}
p(z,m,t)=z^{2m}e^{-z^2\tau}p_0(z)/\tilde{F}^2,
\end{eqnarray}
where we introduced the dimensionless time $\tau=2|C|^2\kappa t$ and
new normalization coefficient $\tilde{F}$ such that $\sum_{z=0}^N
p(z,m,t)=1$.

If the initial atom number $z$ at $K$ sites is uncertain, $p_0(z)$
is broad. For the SF state the probability to find the atom number
$z$ at the lattice region of $K$ sites is given by the binomial
distribution \cite{NP}
\begin{eqnarray}\label{16}
p_\text{SF}(z)=\frac{N!}{z!(N-z)!}\left(\frac{K}{M}\right)^z
\left(1-\frac{K}{M}\right)^{N-z}.
\end{eqnarray}
For a lattice with the large atom and site numbers $N,M\gg1$, but
finite $N/M$, it can be approximated as a Gaussian distribution
\begin{eqnarray}\label{17}
p_\text{SF}(z)=\frac{1}{\sqrt{2\pi}\sigma_z}e^{-\frac{(z-z_0)^2}{2\sigma_z^2}}
\end{eqnarray}
with the mean atom number $\langle \hat{N}_K\rangle = z_0=NK/M$ and
$\sigma_z=\sqrt{N(K/M)(1-K/M)}$ giving the full width at a half
maximum (FWHM) $2\sigma_z\sqrt{2\ln 2}$. The atom number variance in
the SF state is $(\Delta N_K)^2=\langle \hat{N}_K^2\rangle - \langle
\hat{N}_K\rangle^2=\sigma_z^2$.

Eq.~(\ref{15}) shows how the initial distribution $p_0(z)$ changes
in time. The function $z^{2m}\exp{(-z^2\tau)}$ has its maximum at
$z_1=\sqrt{m/\tau}$ and the FWHM $\delta z \approx
\sqrt{2\ln2/\tau}$ (for $\delta z \ll z_1$). Thus, multiplying
$p_0(z)$ by this function will shrink the distribution $p(z,m,t)$ to
a narrow peak at $z_1$ with the width decreasing in time.

Physically, this describes the projection of the atomic quantum
state to a final state with the squeezed atom number at $K$ sites (a
Fock states $|z_1,N-z_1\rangle$ with the precisely known atoms at
$K$ sites $z_1$ and $N-z_1$ atoms at $M-K$ sites).  Thus, the final
quantum state of the light-matter system is
\begin{eqnarray}\label{18}
|\Psi_c\rangle=|z_1,N-z_1\rangle|\alpha_{z_1}\rangle,
\end{eqnarray}
which is a product state showing that the light and matter get
disentangled. {\it A priori} $z_1$ is unpredictable. However,
measuring the photon number $m$ and time $t$, one can determine
$z_1$ of the quantum trajectory.

Even the final Fock state (\ref{18}) can contain the significant
atom-atom entanglement, as this is still a many-body state. In
general, the Fock state $|z_1,N-z_1\rangle$ cannot be factorized
into the product of two states for $K$ illuminated and $M-K$
unilluminated lattice sites. Thus, the entanglement can survive even
between two lattice regions, which depends on the value of $z_1$
realized at a particular quantum trajectory. For some cases, the
factorization is possible. For example (cf. Ref. \cite{PRL09}), the
initial superfluid state after the measurement approaches the
product of two superfluids:
$|z_1,N-z_1\rangle=|SF\rangle_{z_1,K}|SF\rangle_{N-z_1,M-K}$.

Note, that our model does not specify how $K$ sites were selected,
which is determined by the lattice and light geometry. The simplest
case is to illuminate a continuous region. However, one can also
illuminate each second site by choosing the probe wavelength twice
as lattice period and get an atom number squeezing at odd and even
sites. In this way, one gets a measurement-prepared product of two
SFs ``loaded'' at sites one by one (e.g. atoms at odd sites belong
to one SF, while at even sites to another). While the initial SF, as
usual, shows the long-range coherence $\langle b^\dag_i b_j\rangle$
with the lattice period, the measurement-prepared state will
demonstrate the doubled period in $\langle b^\dag_i b_j\rangle$
($b_j$ is the atom annihilation operator such that
$b^\dag_jb_j=\hat{n}_j$). Thus, even though our model does not
include the matter filed operators $b_j$, but only the atom number
operators $\hat{n}_j$, the matter coherence can be still affected
and modified by our QND measurement scheme in a nontrivial way.

The conditioned cavity photon number $\langle
a^\dag_1a_1\rangle_c(t)=|C|^2 \sum_{z=0}^Nz^2p(z,m,t)$ is given by
the second moment of $p(z,m,t)$. Finally, it reduces to $\langle
a^\dag_1a_1\rangle_c=|C|^2z_1^2$, reflecting a direct correspondence
between the final atom number and cavity photon number, which is
useful for the experimental measurements.

Further details of the light measurement at a diffraction maximum
were presented in Ref.~\cite{PRL09}.

\subsection{Preparation of the Schr{\"o}dinger cat states}

The macroscopic superposition state (Schr{\"o}dinger cat state) can
be prepared detecting light at the direction of a diffraction
minimum \cite{PRL09}. The condition of the diffraction minimum for
the scattering of light from the probe wave $a_0$ into the cavity
mode $a_1$ is the following: the atoms at the neighboring lattice
sites scatter the light with the phase difference $\pi$. For the
plain standing or traveling waves, this condition means that in the
expression for the operator $\hat{D}_{10}= \sum_{j=1}^K{u_1^*({\bf
r}_j)u_0({\bf r}_j)\hat{n}_j}$, $u_1^*({\bf r}_j)u_0({\bf
r}_j)=(-1)^{j+1}$. Thus, $\hat{D}_{10}=\sum_{j=1}^M
(-1)^{j+1}\hat{n}_j$ is the operator of atom number difference
between odd and even sites (in this subsection, we consider all
sites illuminated, $K=M$). Similarly to the previous subsection, in
Eq.~(\ref{7}), after neglecting the dispersive frequency shift, the
only statistical quantity is $D^q_{10}$ giving the atom number
difference for the configuration $q$. We will call this single
statistical quantity as $z$: $D^q_{10}=z$, which varies between $-N$
and $N$ with a step 2 reflecting all possible realizations of the
atom number difference.

Equations (\ref{14}) and (\ref{15}) keep their form for the
diffraction minimum as well, however, with a different meaning of
the statistical variable $z$, which is now a realization of the atom
number difference, and $p(z,m,t)$ is its probability.

For the SF state the probability to find the atom number at odd (or
even) sites $\tilde{z}$ [$\tilde{z}=(z+N)/2$ because the atom number
difference is $z$ and the total atom number is $N$] is given by the
binomial distribution \cite{NP}
\begin{eqnarray}\label{19}
p_\text{SF}(\tilde{z})=\frac{N!}{\tilde{z}!(N-\tilde{z})!}
\left(\frac{Q}{M}\right)^{\tilde{z}}
 \left(1-\frac{Q}{M}\right)^{N-\tilde{z}},
\end{eqnarray}
where $Q$ is the number of odd (or even) sites. For even $M$,
$Q=M/2$ and Eq.~(\ref{19}) simplifies. For a lattice with the large
atom and site numbers $N,M\gg1$, but finite $N/M$, this binomial
distribution, similarly to the previous subsection, can be
approximated by a Gaussian function. Changing the variable as
$z=2\tilde{z}-N$ we obtain the Gaussian function for the probability
to find the atom number difference $z$:
\begin{eqnarray}\label{20}
p_\text{SF}(z)=\frac{1}{\sqrt{2\pi}\sigma_z}e^{-\frac{z^2}{2\sigma_z^2}}
\end{eqnarray}
with the zero mean $z$ and $\sigma_z=\sqrt{N}$ giving the FWHM
$2\sigma_z\sqrt{2\ln 2}$. The variance of the atom number difference
in the SF state is $\sigma_z^2=N$.

The striking difference from the diffraction maximum is that our
measurement and the probability (\ref{15}) are not sensitive to the
sign of $z$, while the amplitudes $\alpha_z=Cz$ are. So, the final
state obtained from Eq.~(\ref{7}) is a superposition of two Fock
states with $z_{1,2}=\pm \sqrt{m/\tau}$ and different light
amplitudes: $\alpha_{z_2}=-\alpha_{z_1}$,
\begin{eqnarray}\label{21}
|\Psi_c\rangle=\frac{1}{\sqrt{2}}(|z_1\rangle|\alpha_{z_1}\rangle+
(-1)^m|-z_1\rangle|-\alpha_{z_1}\rangle).
\end{eqnarray}

In contrast to a maximum, even in the final state, the light and
matter are not disentangled. In principle, one can disentangle light
and matter by switching off the probe and counting all leaking
photons. Then both $|\alpha_{z_1}\rangle$ and
$|-\alpha_{z_1}\rangle$ will go to the vacuum $|0\rangle$. Thus one
can prepare a quantum superposition of two macroscopic atomic states
$(|z_1\rangle+(-1)^m|-z_1\rangle)/\sqrt{2}$, which is a
Schr{\"o}dinger cat state that, in the notation of odd and even
sites, reads
\begin{eqnarray}
\frac{1}{\sqrt{2}}\left(|\frac{N+z_1}{2},\frac{N-z_1}{2}\rangle +
(-1)^m|\frac{N-z_1}{2},\frac{N+z_1}{2}\rangle \right). \nonumber
\end{eqnarray}

We have shown that the detection of photons at the direction of a
diffraction minimum leads to the preparation of the Schr{\"o}dinger
cat state (\ref{21}). The physical reason for this is that the
quantum measurement of photons determines the absolute value of the
atom number difference $|z|$. However, as the photon number is not
sensitive to the sign of $z$, one ends up in the superposition of
states with the positive and negative values of $z$.

Unfortunately, Eq.~(\ref{21}) demonstrates a very strong
disadvantage of such a method, which makes it difficult to realize
experimentally. Each photodetection flips the sign between two
components of the quantum superposition in Eq.~(\ref{21}). This
means, that if one loses even a single photocount, which is very
probable for a realistic photodetector, one ends up in the mixture
of two state (\ref{21}) with plus and minus signs. Such a mixed
state does not contain any atomic entanglement any more, in contrast
to the pure state (\ref{21}), which is a highly entangled one.

Formally, the appearance of the sign flip $(-1)^m$ in Eq.~(\ref{21})
originates from Eq.~(\ref{7}), which contains the coefficient
$\alpha_z^m$ in the pre-factor of each Fock state. Since, in the
final state, two components with opposite signs of light amplitudes
survive ($\alpha_{z_2}=-\alpha_{z_1}$), the term $\alpha_z^m$
produces the coefficient $(-1)^m$. Therefore, even one
photodetection changes the phase between two components in
Eq.~(\ref{21}) by $\Delta \varphi_1=\pi$, which is the maximal
possible phase difference. The idea to make the preparation scheme
more stable with respect to the photon losses is based on the
possibility to make this phase jump $\Delta \varphi_1$ smaller than
$\pi$. In this case, loosing one photon will also lead to the
mixture of two cat states. However, if those cat states are not very
different (i.e. the phase difference $\Delta \varphi_1$ is small),
the mixed state will still contain significant atomic entanglement.

In Sec. VI, we present a scheme to prepare the Schr{\"o}dinger cat
state with a phase difference between two components smaller than
$\pi$. Thus, such a scheme is more practical and robust with respect
to the photon losses and, hence, decoherence.

The atom number squeezed states (\ref{18}) prepared by observing
light at a diffraction maximum is indeed more robust than the
Schr{\"o}dinger cat state (\ref{21}) obtained at a diffraction
minimum, as the former do not have any phase jump. However, the
convenient property of the measurement at a minimum is that during
the same time interval (e.g., during the shrinking time which is the
same for a maximum and minimum, $\delta z \approx
\sqrt{2\ln2/\tau}$) the number of photons scattered at a diffraction
minimum ($\langle a_1^\dag a_1\rangle =|C|^2N$) is much smaller than
the one scattered at a maximum ($\langle a_1^\dag a_1\rangle
=|C|^2N_K^2$) \cite{PRL07,PRA07,PRL09}.

Further details of the light measurement at a diffraction minimum
were presented in Ref.~\cite{PRL09}.

\section{Photon statistics}

In this section, we consider three kinds of photon statistics: (i)
statistics $p_\Phi(n,m,t)$ of the photon number $n$ in a cavity
after $m$ photons were detected outside the cavity, (ii) statistics
$P(m,t)$ of the photocount number $m$, and (iii) statistics
$\tilde{P}_T(m,t)$ of the photocount number $m$ if, after the time
measurement $T$, $m_T$ photons were detected.

First, let us consider the statistics of the number of photons in a
cavity after the measurement time $t$ and $m$ photodetections. The
joint probability to find a number of photons in a cavity $n$
together with finding the atomic state in the Fock state
$|q\rangle=|q_1,..,q_M\rangle$ is obtained by projecting the general
solution (\ref{6}) on the state $|q\rangle|n\rangle$ and is given by
\begin{eqnarray}
W_q(n,m,t)=\frac{|\alpha_q(t)|^{2n}}{n!}e^{-|\alpha_q(t)|^2}p_q(m,t),\nonumber
\end{eqnarray}
where the probability to find the atomic Fock state $p_q(m,t)$ is
given by Eq. (\ref{10}). However, the probability to find $n$
photons in a cavity independently of the atomic state is obtained by
projecting the solution (\ref{6}) on the light Fock state
$|n\rangle$ and taking the trace over the atomic states. Thus, the
cavity photon number distribution function is given by
\begin{eqnarray}
p_{\Phi}(n,m,t)=\sum_q\frac{|\alpha_q(t)|^{2n}}{n!}e^{-|\alpha_q(t)|^2}p_q(m,t),
\nonumber
\end{eqnarray}
where the sum is taken over all possible configurations $q$.

If, as in the previous section, the only atomic statistical quantity
is $z$, the sum is simplified and the probability $p_{\Phi}(n,m,t)$
to find $n$ photons in a cavity after the measurement time $t$ and
$m$ photodetections reads as
\begin{eqnarray}\label{1x}
p_{\Phi}(n,m,t)=\sum_z\frac{|\alpha_z(t)|^{2n}}{n!}e^{-|\alpha_z(t)|^2}p(z,m,t),
\end{eqnarray}
where the atom number distribution $p(z,m,t)$ is given by
Eq.~(\ref{12}).

In general, $p_{\Phi}(n,m,t)$ is a super-Poissonian distribution.
During the measurement, the atomic distribution $p(z,m,t)$ shrinks
to one or two symmetric peaks corresponding to the atom-number
squeezed or macroscopic superposition states. Thus, after some time,
only a single term (or two equal terms) survives in the sum, and the
cavity photon statistics $p_{\Phi}(n,m,t)$ evolves from
super-Poissonian to Poissonian one. This fact can be checked
experimentally.

During the measurement, the mean conditional photon number $\langle
a_1^\dag a_1\rangle_c$ approaches the value $|\alpha_{z_1}|^2$,
which enables one to determine the final atom number $z_1$ by
measuring the photon number in a cavity.

Let us now consider the probability $P(m,t)$ to detect $m$ photons
within the time $t$, if the initial atom number distribution is
$p_0(z)$. As shown, for example, in Ref.~\cite{Ueda}, this
probability can be obtained from the state (\ref{6}) using the
integration over all detection moments $t_1,t_2,..,t_m$ from 0 to
$t$ (because we are not interested in the time moments, but only in
the total number of the photocounts $m$) and taking the trace. The
simple result can be obtained for the case $t,t_1>1/k$, since the
solution (\ref{7}) does not depend on the detection times. The
probability reads
\begin{eqnarray}\label{2x}
P(m,t)=\sum_z\frac{(2\kappa|\alpha_z|^2t)^m}{m!}e^{-2\kappa|\alpha_z|^2t}p_0(z),
\end{eqnarray}
where the powers of $m$ appear due to the $m$ time-integrations.

In contrast to the probability $p_{\Phi}(n,m,t)$, Eq.~(\ref{1x}),
which characterizes the conditional distribution of the cavity
photons at a particular trajectory, the probability $P(m,t)$ depends
on the initial atom number distribution $p_0(z)$ and is not a
characteristic of a particular quantum trajectory, but rather of an
ensemble average. In general, this is a super-Poissonian
distribution, which does not approach any Poissonian one. From
Eq.~(\ref{2x}), the increase of the mean photocount number with time
is given by
\begin{eqnarray}
\langle m\rangle_0 = 2\kappa t \langle a_1^\dag
a_1\rangle_0,\nonumber
\end{eqnarray}
where $\langle a_1^\dag a_1\rangle_0$ is not a conditional photon
number, but the one calculated for the initial atomic state. In
average, the photocount number linearly increases in time. However,
as the distribution is not Poissonian, the fluctuations of the
photocount rate $m/t$ do not decrease to zero.

One can also introduce another statistical distribution: the
distribution $\tilde{P}_T(m,t)$ of the photocount number $m$ if,
after the measurement time $T$, $m_T$ photons have been already
detected. Similar approach as in Eq.~(\ref{2x}), leads to the
following result:
\begin{eqnarray}\label{3x}
\tilde{P}_T(m,t)=\sum_z\frac{[2\kappa|\alpha_z|^2(t-T)]^m}{m!}e^{-2\kappa|\alpha_z|^2(t-T)}
\nonumber\\
\times p(z,m_T,T),
\end{eqnarray}
which, in contrast to Eq.~(\ref{2x}), depends not on the initial
atomic distribution, but on that at time $T$. Thus, this probability
combines the quantum trajectory evolution up to the time $T$, and
the ensemble average after that. As we know, the atomic distribution
$p(z,m_T,T)$ approaches the single peak for the Fock state with
increasing $T$. Therefore, this photocount probability approaches
the Poissonian distribution with increasing $T$, the fluctuations of
$m$ grows in time as $\sqrt{m}\sim \sqrt{t}$, and the fluctuations
of the photocount rate $m/t$ vanishes with increasing time.

\section{State preparation by cavity transmission measurement}

We now switch to a different probing scheme, using the probe through
the mirror with the amplitude $\eta$ (Fig. 1), and assuming no
transverse probe, $a_0=0$. From Eq.~(\ref{8}) we see that the light
amplitudes depend only on the single statistical quantity $D^q_{11}=
\sum_{j=1}^K{|u_1({\bf r}_j)|^2q_j}$, which reduces to the atom
number at $K$ sites for the traveling wave $a_1$ at any angle to the
lattice, or for the standing wave $a_1$ with atoms trapped at the
antinodes. Thus, in this case, the statistical quantity is
$z=D^q_{11}$, which changes between 0 and $N$. The term
$U_{11}D^q_{11}$ in Eq.~(\ref{8}) has a meaning of the dispersive
frequency shift of the cavity mode, due to the presence of atoms in
a cavity.

The light amplitude $\alpha_z$ from Eq.~(\ref{8}) can be rewritten
as a function of the atom number at $K$ sites $z$:
\begin{eqnarray}\label{22}
\alpha_z=C'\frac{\kappa/U_{11}}{i(z-z_p)+\kappa/U_{11}} \text{ with
} C'=\frac{\eta}{\kappa},
\end{eqnarray}
and the parameter $z_p=\Delta_p/U_{11}$ is fixed by the probe-cavity
detuning $\Delta_p$. The probability to find the atom number $z$
(\ref{13}) takes the form
\begin{eqnarray}\label{23}
p(z,m,t)=\frac{e^{-\tau'(\kappa/U_{11})^2/[(z-z_p)^2+(\kappa/U_{11})^2]}}
{[(z-z_p)^2+(\kappa/U_{11})^2]^m} p_0(z)/F'^2,
\end{eqnarray}
where the dimensionless time is $\tau'=2|C'|^2\kappa t$.

The pre-factor function in front of $p_0(z)$ in Eq.~(\ref{23}) has a
form more complicated than the one we had for the transverse probing
in Eq.~(\ref{15}). However, it provides us a richer physical
picture. In contrast to the transverse probing, this function allows
us a collapse to both the singlet and doublet distribution. Thus,
the measurement of the cavity transmission contains both cases of
transverse probing at a diffraction maximum and minimum considered
in Sec. IV.

\begin{figure*}
\scalebox{0.9}[0.9]{\includegraphics{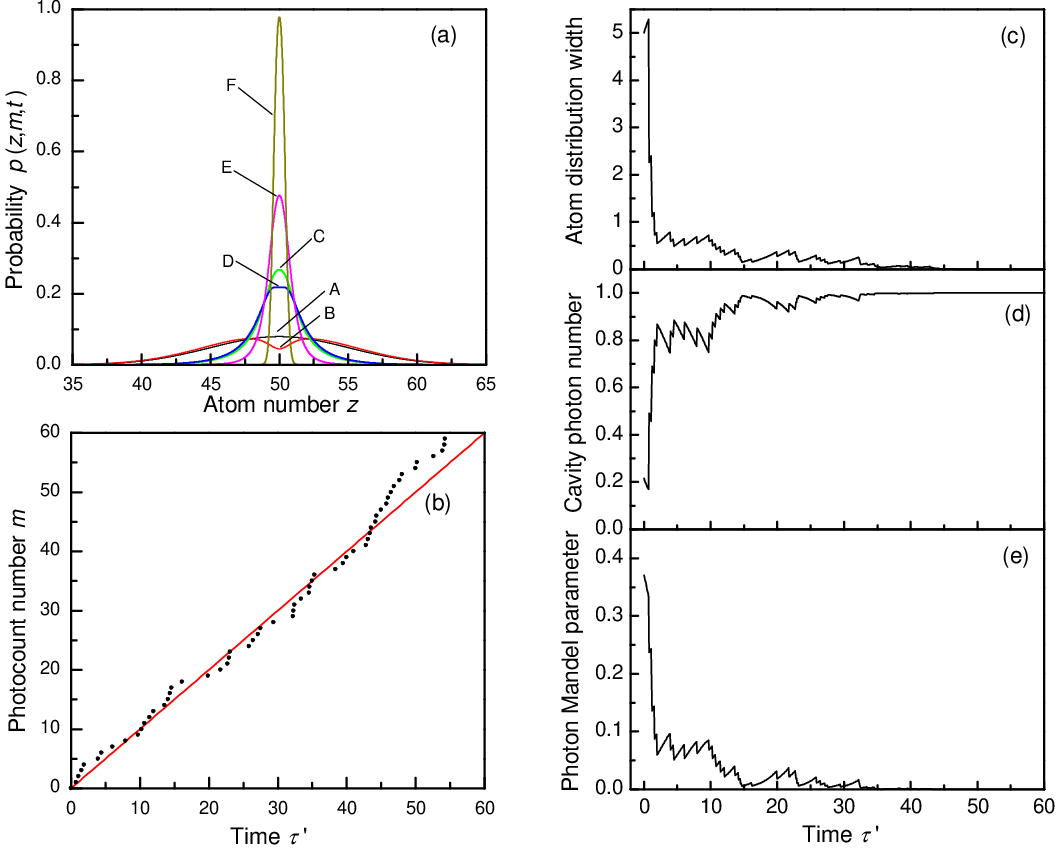}}
\caption{\label{fig2}(Color online) Photodetection trajectory
leading to a single-peak distribution (atom number squeezing). The
probe-cavity detuning is chosen such that $z_p=50$ coincides with
the initial distribution center. (a) Shrinking atom number
distribution at different times $\tau'=$ 0 (number of photocounts is
$m=0$), 0.7 (just before the first count, $m=0$), 0.7 (just after
the first count, $m=1$), 1.1 ($m=1$), 1.1 ($m=2$), 14.6 ($m=17$)
(A-F); (b) time dependence of the photocount number $m$, the red
line is for $m=\tau'$; (c) decreasing width of the atom number
distribution; (d) reduced conditioned photon number $\langle
a^\dag_1 a_1\rangle_c/|C'|^2$ with quantum jumps; (e) photon Mandel
parameter. Initial state: SF, $N=100$ atoms, $M=100$ lattice sites,
$K=M/2=50$ illuminated sites.}
\end{figure*}

If for a particular quantum trajectory the number of photocounts is
large such that $m/\tau'\ge 1$, the distribution $p(z,m,t)$
collapses to a single-peak function at $z_p$. This singlet shrinks
(FWHM) as $\delta z\approx 2(\kappa/U_{11})\sqrt[4]{2\ln{2}/\tau'}$,
which can be very fast for a high-Q cavity with the small $\kappa$.
In the estimation of $\delta z$, we used the assumption that the
peak has already become narrow, $\delta z \ll \kappa/U_{11}$, and
that $m\approx 2\kappa\langle a_1^\dag a_1\rangle_c t$. Similar to
the transverse probing, in the final state, the light and atoms are
disentangled: $|z_p,N-z_p\rangle|\alpha_{z_p}\rangle$, which
corresponds to the atom-number squeezed (Fock) state and coherent
light state.

If, however, the number of photocounts is small, $m/\tau'< 1$, the
distribution $p(z,m,t)$ collapses to a doublet centered at $z_p$
with two satellites at $z_{1,2}=z_p\pm\Delta z$ with
\begin{eqnarray}\label{24}
\Delta z = \frac{\kappa}{U_{11}}\sqrt{\frac{\tau'}{m}-1}.
\end{eqnarray}
When the doublet has become well-separated (i.e. $\delta z \ll
\Delta z$), each of its component shrinks in time as
\begin{eqnarray}
\delta z \approx \Delta z \left(1+\frac{\kappa^2}{U^2_{11}\Delta
z^2}\right) \sqrt{\frac{2\ln{2}}{\tau'}\left(1+\frac{U^2_{11}\Delta
z^2}{\kappa^2}\right)}. \nonumber
\end{eqnarray}

Physically, tuning the probe at $\Delta_p$, we may expect scattering
from the atom number $z_p$ providing such a frequency shift
$\Delta_p$. If the photocount number is large ($m/\tau'\ge1$),
indeed, the atom number is around $z_p$ and it collapses to this
value. However, if $m$ is small, we gain knowledge that the atom
number $z$ is inconsistent with this choice of $\Delta_p$, but two
possibilities $z<z_p$ or $z>z_p$ are indistinguishable. This
collapses the state to a superposition of two Fock states with
$z_{1,2}$, symmetrically placed around $z_p$.

Thus, the transmission measurement scheme allows one to prepare both
the atom number squeezed state and the Schr{\"o}dinger cat state.
The appearance of the singlet for squeezed state or the doublet for
cat state can be determined by measuring $m$ and $t$, or the final
photon number $\langle a^\dag_1a_1\rangle_c=|\alpha_{z_1}|^2$. From
Eq.~(\ref{22}), we see that there is a direct correspondence between
the final cavity photon number and the parameter of the doublet
$\Delta z$:
\begin{eqnarray}\label{xx}
\langle a^\dag_1a_1\rangle_c = \frac{\eta^2}{U^2_{11}\Delta
z^2+\kappa^2},
\end{eqnarray}
which makes the experimental determination of the doublet position
possible. The parameter $\Delta z$ can be also determined by
measuring the photocount number $m$ and time $t$ using
Eq.~(\ref{24}).

In Figs. 2-5, we present the results, where the quantum trajectories
of qualitatively different kinds were realized. In all figures, the
initial state is the superfluid with the atom number $N=100$ at
$M=100$ lattice sites, the half of all lattice sites $K=50$ are
illuminated by the cavity mode. The initial distribution of the atom
number at $K$ sites is given by Eq.~(\ref{16}) and can be well
approximated by the Gaussian distribution (\ref{17}) with the mean
value $\langle \hat{N}_K\rangle = z_0=NK/M=50$ and
$\sigma_z=\sqrt{N(K/M)(1-K/M)}=5$.

\begin{figure*}
\scalebox{0.9}[0.9]{\includegraphics{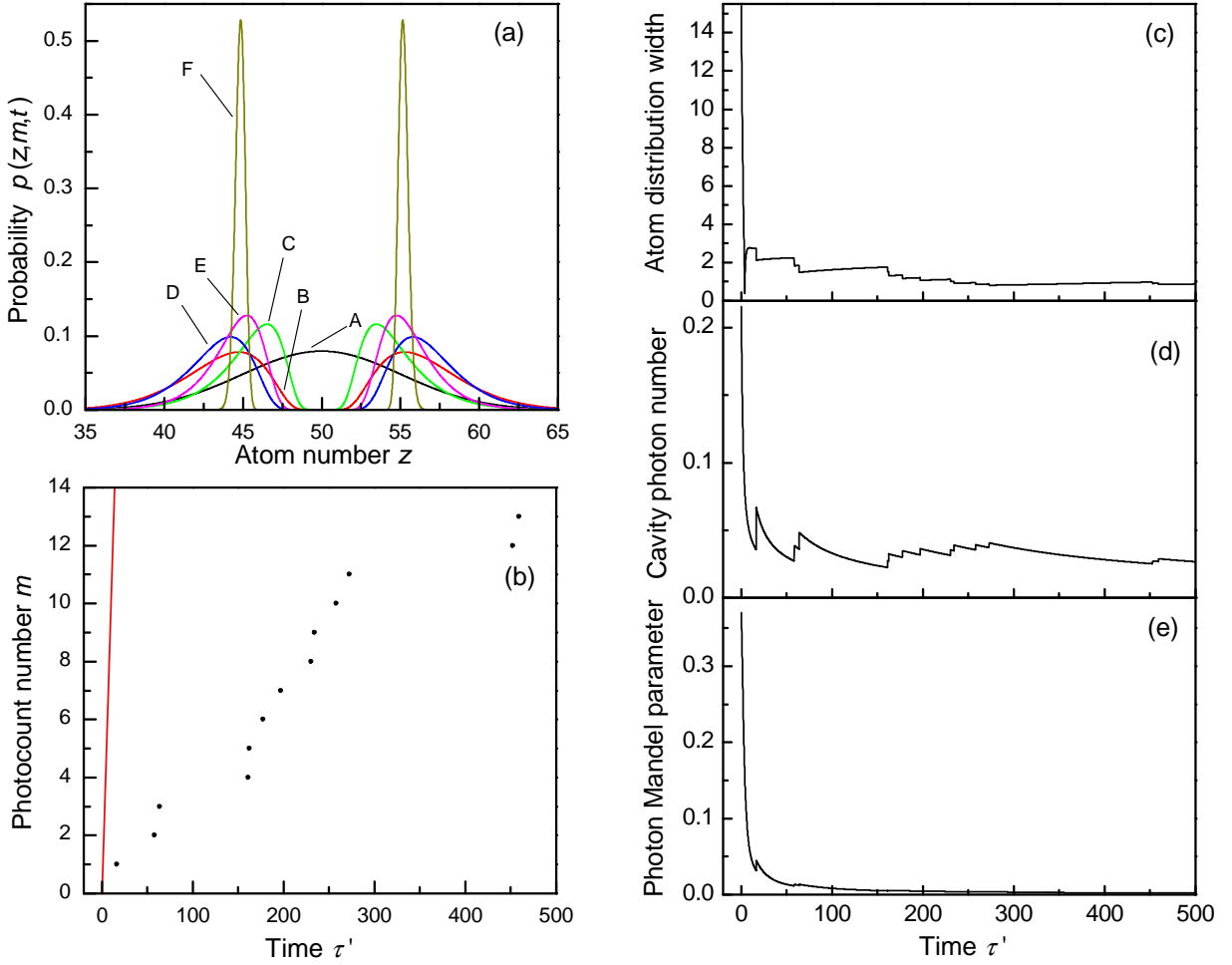}}
\caption{\label{fig3}(Color online) Photodetection trajectory
leading to a doublet distribution (Schr{\"o}dinger cat state). The
probe-cavity detuning is chosen such that $z_p=50$ coincides with
the initial distribution center. (a) Shrinking atom number
distribution at different times $\tau'=$ 0 (number of photocounts is
$m=0$), 16.4 (just before the first count, $m=0$), 16.4 (just after
the first count, $m=1$), 58.2 ($m=1$), 58.2 ($m=2$), 2017.6 ($m=73$)
(A-F); (b) time dependence of the photocount number $m$, the red
line is for $m=\tau'$; (c) decreasing width of one of the peaks in
the atom number distribution; (d) reduced conditioned photon number
$\langle a^\dag_1 a_1\rangle_c/|C'|^2$ with quantum jumps; (e)
photon Mandel parameter. Initial state: SF, $N=100$ atoms, $M=100$
lattice sites, $K=M/2=50$ illuminated sites.}
\end{figure*}

Figure 2 presents the results for a quantum trajectory, which leads
to the collapse to a single-peak distribution. In this case, the
probe is detuned in such way, that the probe-cavity detuning
corresponds to the center of the atom number distribution:
$z_p=\Delta_p/U_{11}=50$. In Fig. 2(a), the evolution of the atom
number probability $p(z,m,t)$, Eq.~(\ref{23}), is shown. The curve A
is the initial atom number distribution having the Gaussian shape.
Curve B shows the non-Hermitian evolution of the probability just
before the first jump ($m=0$). As, at this time interval, the number
of photocounts is small ($m/\tau'<1$), the distribution tends to a
doublet. This is assured by the exponential factor in Eq.~(\ref{23})
leading to the suppression of the component at $z=z_p$ as one does
not record the photocounts at the expected detuning. However, just
after the photocount ($m=1$) this distribution instantly changes to
the curve C, which has already a single peak as for this trajectory
it turns out that $m/\tau'>1$. The switch to a single peak is
assured by the Lorentzian factor in Eq.~(\ref{23}). After that,
before the second jump, the probability decreases at $z=z_p$ and
broadens (curve D), but jumps upwards again and narrows, when the
second jump occurs (curve E). Finally, after many jumps, the
probability distribution becomes narrow and has a single peak at
$z=z_p$ (curve F).

Figure 2(b) shows the number of photocounts $m$ growing in time. It
is clear that $m$ stays always near the line $m=\tau'$. The
appearance of the singlet is assured by the fact, that at the
initial stage $m/\tau'>1$.

Figure 2(c) shows the evolution of the width of the atom number
distribution, $\sqrt{(\Delta N_K)^2}$, which decreases to zero
reflecting the shrinking distribution. Figure 2(d) shows the reduced
conditioned photon number in the cavity $\langle a_1^\dag
a_1\rangle_c/|C'|^2$. One sees that for the initial atom
distribution it starts from a relatively small value. However, as
the atomic state goes to the Fock state with atom number $z_p$
exactly corresponding to the detuning $\Delta_p$, it approaches the
maximal possible value $\langle a_1^\dag a_1\rangle_c/|C'|^2=1$.
Moreover, one can easily see the quantum jumps in the initial stage,
when the light field consists of several coherent-state components.
Finally, the jumps disappear as the light state approaches a single
coherent state $|\alpha_{z_p}\rangle$, when the atomic state
approaches a Fock state. It is interesting to note, that the photon
escape from the cavity (photocount) leads to the increase of the
conditional photon number in the cavity, while the no-count process
leads to its decrease. This is a counter-intuitive characteristic
feature of the super-Poissonian photon statistics and is determined
by the conditional nature of the probabilities considered
\cite{Ueda}. Figure 2(e) shows the reduced Mandel parameter
$Q/|C'|^2$ characterizing the photon number variance: $Q=(\langle
n_{\Phi}^2\rangle-\langle n_{\Phi}\rangle^2)/\langle
n_{\Phi}\rangle-1$, where $n_{\Phi}=\langle a_1^\dag a_1\rangle_c$
is the conditioned photon number. One can see how it decreases to
zero corresponding to the coherent state of light. The parameters
shown in Figs. 2(b)-2(e) can be measured experimentally thus
presenting the verification of our theory.

Figure 3 presents the results for the case, where the detuning also
corresponds to the atom number distribution center $z_p=50$, but the
quantum trajectory leads to the doublet distribution (Sch{\"o}dinger
cat state). Figure 3(a) shows the evolution of the atom number
distribution $p(z,m,t)$. Curve A is the initial Gaussian
distribution. Before the first jump (curve B) the distribution
evolves to a doublet-like, similar to the curve B in Fig. 2(a).
However, in contrast to Fig. 2(a), the first photocount does not
return the distribution back to the single peak, and it stays
doublet-like (curve C). This is so, because the first jump occurs
rather late such that $m/\tau'<1$. Next jumps are rather late as
well, so the distribution evolves to a doublet (curves D, E, and F).

Figure 3(b) shows the number of photocounts $m$ growing in time. It
is clear that, in contrast to Fig. 2(b), $m$ is always much smaller
than the line $m=\tau'$ (red line), which gives the experimental
possibility to claim the appearance of the doublet.

Figures 3(c), 3(d), and 3(e), similarly to Fig. 2, show the decrease
of the width of one of two peaks of the atom number distribution,
conditioned cavity photon number with disappearing jumps, and the
Mandel parameter approaching zero. Note, that in contrast to Fig.
2(d), the conditioned photon number $\langle a_1^\dag
a_1\rangle_c/|C'|^2$ does not approach the maximal value 1, but
rather decreases to a smaller value given by the doublet splitting
$\Delta z$, Eq.~(\ref{xx}). Thus, the appearance of the doublet can
be characterized by experimentally measuring $m$ and $t$ [Fig. 3(b)]
or the cavity photon number [Fig. 3(d)].

\begin{figure}
\scalebox{0.9}[0.9]{\includegraphics{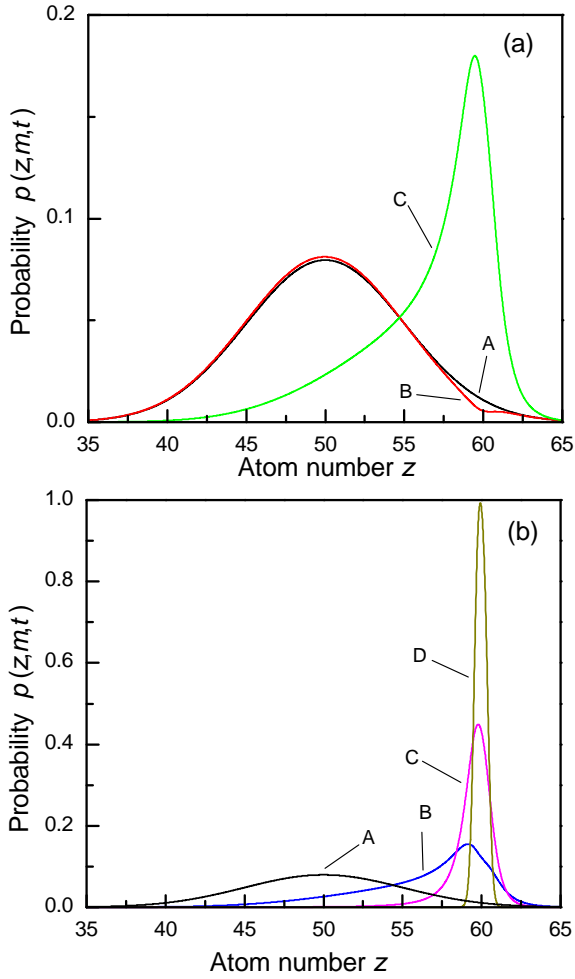}}
\caption{\label{fig4}(Color online) Photodetection trajectory
leading to a single-peak distribution (atom number squeezing). The
probe-cavity detuning is chosen such that $z_p=60$ is at the wing of
the initial distribution. Shrinking atom number distribution at
different times, (a) $\tau'=$ 0 (number of photocounts is $m=0$),
0.7 (just before the first count, $m=0$), 0.7 (just after the first
count, $m=1$) (A-C); (b) $\tau'=$ 0 ($m=0$), 1.1 ($m=1$), 1.1
($m=2$), 14.6 ($m=17$) (A-D). Initial state: SF, $N=100$ atoms,
$M=100$ lattice sites, $K=M/2=50$ illuminated sites.}
\end{figure}

Figures 4 and 5 present another situation, where the probe-cavity
detuning is chosen such that $z_p=60$, which corresponds to a wing
of the atomic distribution function.

Figure 4 shows the evolution of the atom number distribution in
case, where it collapses to a singlet at $z_p$. Thus, here even the
conditioned mean atom number changes from 50 to $z_p=60$. Other
characteristics of this process look very similar to the ones
presented in Figs. 2(b)-(e).

Figure 5 shows the collapse to a doublet distribution around
$z_p=60$ with $\Delta z=7$. Thus, two satellites in Fig. 5(a) are
placed at $z_1=67$ and $z_2=53$. However, while the satellite at
$z_2=53$ is near the maximum of the initial distribution and is well
seen, the second satellite at $z_1=67$ falls on the far wing and is
practically invisible. As a result, the final distribution looks as
a singlet at $z_2=53$, while the second satellite is very small. The
fact that one has indeed a doublet can be verified by measuring the
photocount number [Fig. 5(b)], which is obviously less than $\tau'$,
or by measuring the cavity photon number [Fig. 5(e)], which is less
than the maximal value 1 and depends on $\Delta z$. The measurement
of the mean atom number [Fig. 5(c)] or width of atomic distribution
[Fig. 5(d)] would not distinguish between the singlet and doublet.

This suggests us a method to prepare the macroscopic superposition
state with unequal amplitudes of two components. Choosing the
detuning $\Delta_p$ such that $z_p$ is not at the center of the
initial atom number distribution, one can expect that one of the
satellites will be very probable near the distribution center, while
another one will fall on its wing.

\begin{figure*}
\scalebox{0.9}[0.9]{\includegraphics{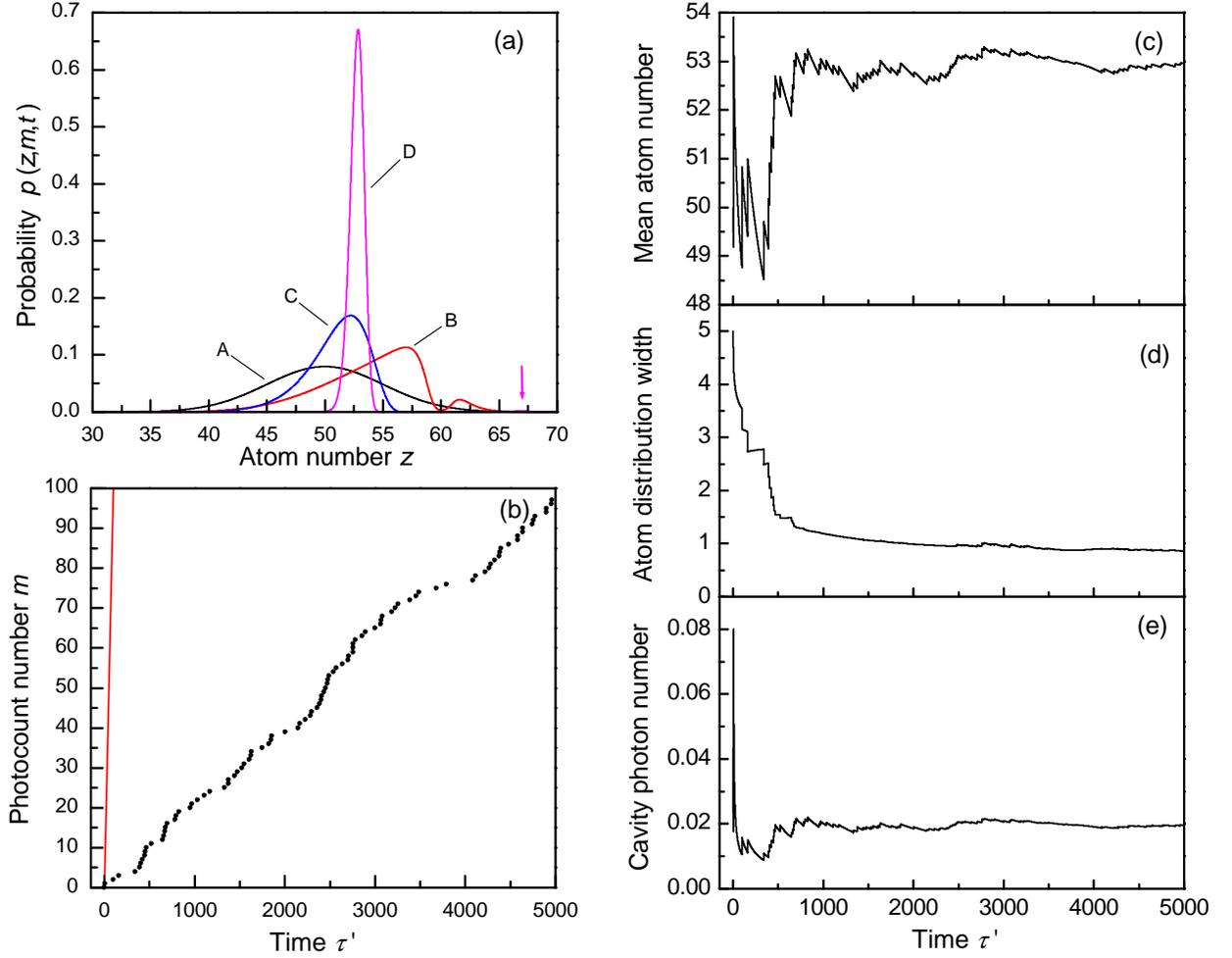}}
\caption{\label{fig5}(Color online) Photodetection trajectory
leading to a doublet distribution (Schr{\"o}dinger cat state). The
probe-cavity detuning is chosen such that $z_p=60$ is at the wing of
the initial distribution. (a) Shrinking atom number distribution at
different times $\tau'=$ 0 (number of photocounts is $m=0$), 5.4
($m=1$), 163.5 ($m=3$), 2006.9 ($m=39$) (A-D); the arrow shows the
position of the very small doublet component, which is almost
invisible; (b) time dependence of the photocount number $m$, the red
line is for $m=\tau'$; (c) conditioned mean atom number; (d)
decreasing width of the atom number distribution; (e) reduced
conditioned photon number $\langle a^\dag_1 a_1\rangle_c/|C'|^2$
with quantum jumps. Initial state: SF, $N=100$ atoms, $M=100$
lattice sites, $K=M/2=50$ illuminated sites.}
\end{figure*}

The most important difference of the Schr{\"o}dinger cat state
prepared by the transmission measurement from the one prepared by
transverse probing (\ref{21}) is that the phase difference between
two components of the quantum superposition is not limited by the
values of $0$ and $\pi$ as in Eq.~(\ref{21}). Although the light
amplitudes corresponding to $z_1$ and $z_2$ has equal absolute
values, their phases are opposite:
$\alpha_{z_1}=|\alpha_{z_1}|\exp{(i\varphi)}$ and
$\alpha_{z_2}=|\alpha_{z_1}|\exp{(-i\varphi)}$, where from
Eq.~(\ref{22}) the light phase $\varphi$ is
\begin{eqnarray}\label{25}
\varphi=-\arctan{\frac{U_{11}\Delta z}{\kappa}}.
\end{eqnarray}
Moreover, the imaginary parts of the functions $\Phi_q$ in
Eqs.~(\ref{7}) and (\ref{9}) are also different for $z_1$ and $z_2$
and have opposite signs: $\Phi(t)=\text{Im}\Phi_{z_1}(t)=
-\text{Im}\Phi_{z_2}(t)=\text{Im}(\eta \alpha_{z_1}^*)$. Using
Eq.~(\ref{22}),
\begin{eqnarray}\label{26}
\Phi(t)=|\alpha_{z_1}|^2 U_{11}\Delta z t.
\end{eqnarray}
Thus, using Eq.~(\ref{7}), the cat state is given by
\begin{eqnarray}\label{27}
|\Psi_c\rangle=\frac{1}{F'}[e^{im\varphi +
i\Phi(t)}|z_1\rangle|\alpha_{z_1}\rangle \sqrt{p_0(z_1)}  \nonumber\\
+ e^{-im\varphi - i\Phi(t)}|z_2\rangle|\alpha_{z_2}\rangle
\sqrt{p_0(z_2)}],
\end{eqnarray}
which is a macroscopic superposition of two Fock states with the
atom numbers $z_1$ and $z_2$ at $K$ sites.

Using the probe-cavity detuning $\Delta_p$ one can chose the center
of the doublet $z_p=\Delta_p/U_{11}$. Moreover, Eq.~(\ref{27}) shows
that using this detuning one can influence, at least,
probabilistically, the ratio between two components in the
superposition (\ref{27}). In particular, if we tune the probe
frequency such that $z_p$ coincides with the center of the initial
atom number distribution $p_0(z)$, the probabilities of the
symmetric doublet components will be equal, $p_0(z_1)=p_0(z_2)$,
providing the equal amplitudes of the cat components in
Eq.~(\ref{27}).

The phases $\Phi(t)$ and $m\varphi$ have opposite evolution in time.
The pase term $\Phi(t)$, Eq.~(\ref{26}), grows in time linearly and
deterministically, while the term $m\varphi$ grows in time
stochastically according to the growth of the photodetection number
$m$. In average, $\langle m\rangle = 2\kappa |\alpha_{z_1}|^2t$.
Thus, in general, the phase difference between two cat components
grows in time linearly. However, for some parameters, the growth of
the average $\langle m\rangle\varphi$ and $\Phi(t)$ can be
compensated. From Eqs.~(\ref{25}) and  (\ref{26}), $\langle
m\rangle\varphi + \Phi(t)=0$ for $U_{11}\Delta z/\kappa\approx
2.33$. So, for the particular cat components, the phase difference
does not grow in average. However, as $m$ is a stochastic quantity,
its uncertainty grows in time as $\sqrt{t}$ (cf. Sec. V). Thus, the
phase difference will still grow in time as $\sqrt{t}$, which is
however much slower than the linear growth.

The problem of photon losses addressed in the previous section is
related to the stochastic quantity $m\varphi$. A single photocount
changes the phase difference between two components by $\Delta
\varphi_1=2\varphi$. In contrast to the state (\ref{21}), where the
phase jump is always maximal $\Delta \varphi_1=\pi$, here, this jump
can be rather small, if the condition $U_{11}\Delta z/\kappa <1$ is
fulfilled. This means, that to provide the robustness of the state
with respect to the photon losses, the doublet should not be split
too strongly. In the next section we quantitatively analyze the
robustness of the cat state.

\section{Robustness of the Schr{\"o}dinger cat states}

The macroscopic superposition state (\ref{27}) is a pure state. In
principle, if the measurement is perfect and all photons $m$ leaking
the cavity are counted by a photodetector, it will evolve according
to Eq.~(\ref{27}) staying pure. However, if one loses one or more
photons, the state becomes a mixture of several states corresponding
to several lost counts $l$. Thus, if $L$ photons are lost, the state
is a mixture of $L+1$ states of the following form for $0<l<L$:
\begin{eqnarray}\label{28}
|\Psi_c\rangle_l=\frac{1}{\sqrt{2}}[e^{il\varphi +
i\gamma}|z_1\rangle|\alpha_{z_1}\rangle + e^{-il\varphi -
i\gamma}|z_2\rangle|\alpha_{z_2}\rangle],
\end{eqnarray}
where, for simplicity, we assumed the symmetric superposition with
$p_0(z_1)=p_0(z_2)$ and included all known phases [for $m$ measured
photons and deterministic $\Phi(t)$] in to the term $\gamma=m\varphi
+ \Phi(t)$. The density matrix of the state (\ref{28}) is
\begin{eqnarray}\label{29}
\rho_l=\frac{1}{2}(|z_1\rangle|\alpha_{z_1}\rangle \langle
z_1|\langle\alpha_{z_1}|+|z_2\rangle|\alpha_{z_2}\rangle \langle
z_2|\langle\alpha_{z_2}| \nonumber \\
+ e^{i2l\varphi + i2\gamma}|z_1\rangle|\alpha_{z_1}\rangle \langle
z_2|\langle\alpha_{z_2}| \nonumber \\
+ e^{-i2l\varphi - i2\gamma}|z_2\rangle|\alpha_{z_2}\rangle \langle
z_1|\langle\alpha_{z_1}|).
\end{eqnarray}
The density matrix of the mixture state describing $L$ lost photons
is given by a sum of the density matrices (\ref{29}):
\begin{eqnarray}\label{30}
\rho^{(L)}=\frac{1}{2}(|z_1\rangle|\alpha_{z_1}\rangle \langle
z_1|\langle\alpha_{z_1}|+|z_2\rangle|\alpha_{z_2}\rangle \langle
z_2|\langle\alpha_{z_2}| \nonumber \\
+ \frac{e^{i2\gamma}}{L+1}\sum_{l=0}^Le^{i2l\varphi}
|z_1\rangle|\alpha_{z_1}\rangle \langle
z_2|\langle\alpha_{z_2}| \nonumber \\
+ \frac{e^{-i2\gamma}}{L+1}\sum_{l=0}^Le^{-i2l\varphi}
|z_2\rangle|\alpha_{z_2}\rangle \langle z_1|\langle\alpha_{z_1}|).
\end{eqnarray}

The quantity characterizing how close is a mixture state to a pure
state is the so-called purity: $P=\text{Tr}(\rho^2)$. For a pure
state it is maximal and equal to 1, while for a maximally mixed
state it is minimal and equal to $1/2$ (in our case of the
two-component states). The purity of the state (\ref{30}) is given
by
\begin{eqnarray}\label{31}
P_L=\frac{1}{2}\left[1+\frac{1}{(L+1)^2}\left|\sum_{l=0}^L
e^{i2l\varphi}\right|^2\right],
\end{eqnarray}
where the sum can be calculated leading to the following result for
the purity of the mixed state corresponding to $L$ lost photons:
\begin{eqnarray}\label{32}
P_L=\frac{1}{2}\left[1+\frac{1}{(L+1)^2}\frac{\sin^2(L+1)\varphi}{\sin^2\varphi}\right].
\end{eqnarray}

For example, in the simplest case, where one photon is lost, the
density matrix of the mixed state is given by the sum of two terms
(\ref{29}) with $l=0$ and $1$:
\begin{eqnarray}\label{33}
\rho^{(1)}=\frac{1}{2}(|z_1\rangle|\alpha_{z_1}\rangle \langle
z_1|\langle\alpha_{z_1}|+|z_2\rangle|\alpha_{z_2}\rangle \langle
z_2|\langle\alpha_{z_2}| \nonumber \\
+ \frac{1}{2}e^{i2\gamma}(1+e^{i2\varphi})
|z_1\rangle|\alpha_{z_1}\rangle \langle
z_2|\langle\alpha_{z_2}| \nonumber \\
+ \frac{1}{2}e^{-i2\gamma}(1+e^{-i2\varphi})
|z_2\rangle|\alpha_{z_2}\rangle \langle z_1|\langle\alpha_{z_1}|),
\end{eqnarray}
which has a purity
\begin{eqnarray}\label{34}
P_1=\frac{1}{2}(1+\cos^2\varphi)=\frac{1}{2}
\left(1+\frac{1}{1+(U_{11}\Delta z/\kappa)^2}\right).
\end{eqnarray}

Equations (\ref{33}) and (\ref{34}) show that if the phase jump
associated with the one-photon lost $\Delta\varphi_1=2\varphi$ is
maximal, $\Delta\varphi_1=\pi$, the state (\ref{33}) is maximally
mixed, because all non-diagonal terms responsible for the quantum
coherence between the states $|z_1\rangle|\alpha_{z_1}\rangle$ and
$|z_2\rangle|\alpha_{z_2}\rangle$ are zero. Its purity (\ref{34}) is
$1/2$, thus no entanglement survived after the single photon lost.
This is a situation of the transverse probing in the diffraction
minimum, Eq.~(\ref{21}), where the phase jump is $\pi$, which makes
the preparation scheme practically difficult.

However, if $\varphi$ is small, the purity (\ref{34}) and the
non-diagonal coefficients in Eq.~(\ref{33}) can be rather large and
close to 1. Thus, after the photon has been lost, one gets a mixed
state, but that of the high purity. More generally, for $L$ photon
losses, Eqs.~(\ref{30}) and (\ref{32}), the total phase jump
$\Delta\varphi_L=2L\varphi$ should be small. Using the expression
for $\varphi$ (\ref{25}), one can estimate the condition for the
high purity of the mixed state as $\varphi<\pi/(2L)$, which for a
small $\varphi$ is approximately the condition $U_{11}\Delta
z/\kappa<\pi/(2L)$. As a result, for the doublet, which is split not
too strongly, the high purity can be preserved even with photon
losses.

\begin{figure}
\scalebox{0.9}[0.9]{\includegraphics{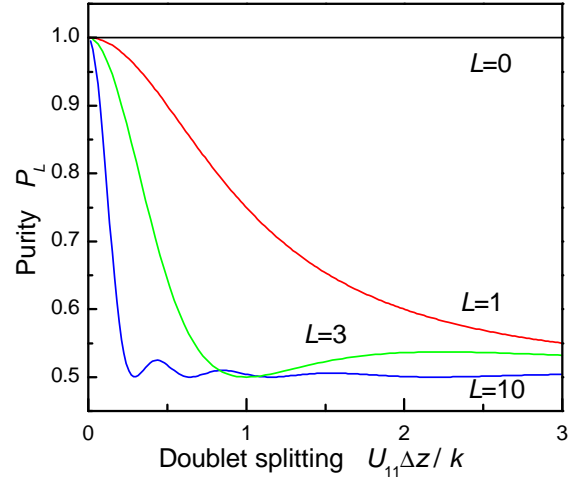}}
\caption{\label{fig6}(Color online) Purity of the mixed state as a
function of the doublet splitting $\Delta z$ for different numbers
$L$ of the photocounts lost: $L=0$, 1, 3, and 10.}
\end{figure}

The purity (\ref{32}) as a function of the doublet splitting $\Delta
z$ for $L=0$, 1, 3, and 10 photon losses is shown in Fig. 6, where
$\varphi$ is given by Eq.~(\ref{25}). One can see the decrease of
the purity with increasing doublet splitting and number of photons
lost. Note the non-monotonous character of its decrease, which means
that the larger splitting does not automatically leads to a smaller
purity.

\section{Conclusions}

We explicitly calculated the quantum state reduction induced by
measurement of off-resonant scattered light from an ultracold
quantum gas trapped in an optical lattice. In our previous papers
\cite{PRL07,NP,PRA07}, the expectation values of various atomic and
light quantities were analyzed, thus, assuming the repeated
measurement to enable the averaging procedure. In contrast, in this
paper, we were focused on a single run of the optical measurement,
i.e., the evolution of the quantities at a single quantum
trajectory. As the scattered light is entangled to the atoms,
quantum back-action of the light measurement alters the atomic
state. The geometry, which determines the form of entanglement, thus
dictates the possible measurement results and final atomic states.

Any quantum state related to an eigenstate of the weighted
atom-number operators $\hat{D}_{lm}$ can thus be prepared in a
probabilistic way. The type of the states can be chosen by the
optical geometry and their probabilities are determined from the
initial distribution. Typically light detection at the diffraction
maximum leads to the preparation of the atom-number squeezed states,
while the detection at a minimum prepares the macroscopic
superposition states peaked at a pair of atom numbers.

The robustness of the resulting Schr{\"o}dinger cat states with
respect to undetected photons was analyzed. The transmission
measurement scheme was shown to prepare the states more robust than
the ones prepared by the detection at a diffraction minimum.

In contrast to recent results in spin squeezing and preparation of
the spin superposition states, which can be also obtained for
thermal atoms
\cite{PolzikNP,PolzikHot,Holland,Genes,MolmerPRA08,MolmerPRA09}, in
our work, quantum nature of ultracold atoms is crucial, as we deal
with the atom number fluctuations appearing due to the
delocalization of ultracold atoms in space.

We demonstrated the time evolution of various measurable quantities
appearing exclusively due to the measurement procedure (as other
obvious sources of dynamics such as tunneling were neglected). The
quantum dynamics is governed by the quantum jumps and conditional
evolution. The quantum state preparation is probabilistic. However,
it can be generalized by including the feedback loop, which will
enable ones the quasi-deterministic state preparation. In this case,
the trapping potential should be continuously modified depending on
the outcome of the photodetector. For example, the detection of
photons at a diffraction maximum squeezes the atomic number at $K$
sites around some value $z_1$, which was not known a priori. The
potential can be continuously tilted in a way to provide the
increase or decrease of this atom number to enable ones to obtained
the number squeezed state with a mean value $\tilde{z}_1$ given a
priori. The same method can be applied for the atom number squeezing
at odd or even sites.

Our results can be applied for other quantum arrays as well, e.g.,
ion strings \cite{ions}, and be useful for the preparation of
various atomic and photonic multipartite entangled states
\cite{Cirac2}. Moreover, they can have a relation to the problem of
interaction of ultracold atoms with other bosonic particles, besides
photons \cite{Jaksch1, Jaksch2}.

Cavity QED with quantum gases can operate with the atom numbers
ranging from millions to one \cite{Rempe}. Thanks to the recent
experimental breakthroughs \cite{Brennecke,Colombe,Slama,Science08},
preparing various kinds of atom number squeezing is already doable,
and creation of, at least, Schr{\"o}dinger ``kittens''
\cite{Grangier,PolzikCat} may become practical.

\begin{acknowledgments}
This work was supported by the Austrian Science Fund FWF (grants
P17709 and S1512).
\end{acknowledgments}

\end{document}